\documentclass[twocolumn,nofootinbib,longbibliography]{revtex4-1}

\usepackage[english]{babel}
\usepackage[toc]{appendix}
\usepackage{natbib}
\usepackage{hyperref}
\usepackage[multiple]{footmisc}
\usepackage[inline]{enumitem}
\usepackage[threshold=1, autopunct, autostyle]{csquotes}

\usepackage{amsmath}
\usepackage{mathrsfs}
\usepackage{amssymb}
\usepackage{amsfonts}
\usepackage{dsfont}
\usepackage{IEEEtrantools}
\usepackage{bm}
\usepackage{amsthm}

\usepackage{xstring}
\usepackage{catchfile}

\makeatletter
\@ifclassloaded{beamer}
  {}
  {
    \bibpunct{[}{]}{,}{n}{}{,}

  }
\makeatother

\renewcommand{\H}{\ensuremath{\mathcal{H}}}
\newcommand{\ket}[1]{|#1 \rangle}

\newcommand{\proj}[1]{|#1\rangle\langle #1|}

\newcommand{\Proj}[1]{\ensuremath{\mathfrak{P}(#1)}}
\newcommand{\ProjH}{\Proj{\H}}

\newcommand{\Z}{\ensuremath{\mathcal{Z}}}

\usepackage{pgfplots}
\usepackage{tikz}
\usetikzlibrary{shapes,calc,positioning,arrows.meta,fadings}

\pgfdeclarelayer{bg}    \pgfsetlayers{bg,main}

\tikzset{colorbox/.style={thick, rounded corners=2pt, text height=1.7ex,text depth=.25ex, draw=#1!70!black, fill=#1!30}}
\tikzset{colorboxS/.style={thick, rounded corners=2pt, text height=1.2ex,text depth=.25ex, draw=#1!70!black, fill=#1!30, font=\footnotesize}}
\tikzset{colorboxXS/.style={thick, rounded corners=2pt, text height=0.7ex,text depth=.10ex, draw=#1!70!black, fill=#1!30, font=\tiny}}
\tikzset{roundedbox/.style={thick, rounded corners=2pt, text height=1.7ex,text depth=.25ex, draw=black}}

\tikzset{CBedgy/.style={thick, text height=1.7ex,text depth=.25ex, draw=#1!70!black, fill=#1!30, font=\ttfamily}}
\tikzset{colorelement/.style={thick, rounded corners=2pt, draw=#1!70!black, fill=#1!30}}

\tikzset{tb/.style={font=\footnotesize, text width=#1, text opacity=1, opacity=.8, fill=white},
  tb/.default=4.5cm}
\tikzset{tbS/.style={font=\scriptsize, text width=#1, text opacity=1, opacity=.8, fill=white},
  tbS/.default=4.5cm}
\tikzset{tbXS/.style={font=\tiny, text width=#1, text opacity=1, opacity=.8, fill=white},
  tbXS/.default=4.5cm}
\tikzset{tbBB/.style={draw=black!70, rounded corners=2pt, font=\footnotesize, text width=#1, text opacity=1, opacity=.8, fill=white},
  tbBB/.default=4.5cm}
\tikzset{tbBBS/.style={draw=black!70, rounded corners=2pt, font=\scriptsize, text width=#1, text opacity=1, opacity=.8, fill=white},
  tbBBS/.default=4.5cm}
\tikzset{tbBBXS/.style={draw=black!70, rounded corners=2pt, font=\tiny, text width=#1, text opacity=1, opacity=.8, fill=white},
  tbBBXS/.default=4.5cm}

\tikzset{quote/.style={thick, draw=black!70, rounded corners=2pt, font=\footnotesize, text width=#1, text opacity=1, opacity=.8, fill=white},
  quote/.default=4.5cm}
\tikzset{quoteS/.style={thick, draw=black!70, rounded corners=2pt, font=\scriptsize, text width=#1, text opacity=1, opacity=.8, fill=white},
  quoteS/.default=4.5cm}
\tikzset{quoteXS/.style={thick, draw=black!70, rounded corners=2pt, font=\tiny, text width=#1, text opacity=1, opacity=.8, fill=white},
  quoteXS/.default=4.5cm}
\tikzset{quoteNB/.style={font=\footnotesize, text width=#1, text opacity=1, opacity=.8, fill=white},
  quoteNB/.default=4.5cm}

\tikzset{comment/.style={thick, draw=black!70, rounded corners=2pt, font=\scriptsize\itshape, text width=#1, text opacity=1, opacity=.8, fill=white},
  comment/.default=4.5cm}
\tikzset{commentS/.style={thick, draw=black!70, rounded corners=2pt, font=\tiny\itshape, text width=#1, text opacity=1, opacity=.8, fill=white},
  commentS/.default=4.5cm}
\tikzset{commentVW/.style={thick, draw=black!70, rounded corners=2pt, font=\scriptsize\itshape, text opacity=1, opacity=.8, fill=white}}
\tikzset{commentSVW/.style={thick, text height=0.7ex,text depth=.10ex, draw=black!70, rounded corners=2pt, font=\tiny\itshape, text opacity=1, opacity=.8, fill=white}}
\tikzset{commentSVWR/.style={thick, text height=0.7ex,text depth=.10ex, draw=black!70, rounded corners=2pt, font=\tiny, text opacity=1, opacity=.8, fill=white}}

\tikzset{conn/.style={thick, shorten <=#1, shorten >=#1}}
\tikzset{tconn/.style={shorten <=#1, shorten >=#1}}
\tikzset{gconn/.style={thick, shorten <=#1, shorten >=#1, draw=gray!80}}

\tikzset{arr_node/.style={pos=0.5,above,font=\scriptsize, sloped}}
\tikzset{ctag/.style={thick, dashed, rounded corners=2pt, text height=1.7ex,text depth=.25ex, draw=#1!70!black, fill=#1!30, font=\ttfamily}}

\tikzset{ist/.style={thick, shorten <=4pt, shorten >=4pt, arrows = {-Bracket[reversed,round]}}}
\tikzset{istgleich/.style={thick, shorten <=4pt, shorten >=4pt, arrows = {Bracket[reversed,round]-Bracket[reversed,round]}}}
\tikzset{nicht/.style={thick, dashed, shorten <=4pt, shorten >=4pt, arrows = {Bracket[round]-}}}

\definecolor{yorange}{HTML}{ff8c00}

\newcommand\blArrow{}
\def\blArrow[#1](#2);
  { \draw[#1] (#2) -- ++(.9,0) -- ++(-.9,-5) -- cycle; }
\newcommand\stdBlArrow{}
\def\stdBlArrow(#1)
  { \blArrow[line width=2pt, draw=red!70!black, fill=red!40, rounded corners=1pt](#1) }

\newcommand\redArrow{}
\def\redArrow(#1);
  { \draw[thick] ($(#1)+(-.3,.1)$) -- ($(#1)+(0,-.1)$) -- ($(#1)+(.3,.1)$);}
\newcommand\fSTS{}
\def\fSTS(#1);
  { \draw[fill=red!40, thick, draw=red!60!black] (#1) circle (.5);
    \draw[->] ($(#1)+(-.05,-.05)$) to ++(0,.4);
    \draw[->] ($(#1)+(-.05,-.05)$) to ++(.4,0);
    \draw[->] ($(#1)+(-.05,-.05)$) to ++(-.2,-.2);}
\newcommand\reduction{}
\def\reduction(#1);
  { \draw[fill=cyan!40, thick, draw=cyan!60!black] (#1) circle (.5);
    \draw[fill=black] ($(#1) + (-.2, .25)$) to ++(0.4,0) to ++(-.3,-.6) to cycle;}
\newcommand\customLegend{}
\def\customLegend(#1);
  {\fSTS(#1);
   \node[right of=#1] {fixed space-time structure};
   \draw[rounded corners=2pt] ($(#1)+(-.5,-.5)$) rectangle ($(#1)+(4, -4)$);}

\makeatletter

\newcommand*{\blitzset}{\pgfqkeys{/blitz}}\blitzset{
  height/.initial=4cm,
  width/.initial=2cm,
  breadth/.initial=.3cm,
  ratio/.initial=.4,
}

\pgfdeclareshape{blitz}
{
  \savedanchor\centerpoint{
    \pgf@x = .5\wd\pgfnodeparttextbox
    \pgf@y = .5\ht\pgfnodeparttextbox
  }
  \anchor{center}{\centerpoint}
  \anchor{north}{\centerpoint\pgfmathsetlength\pgf@ya{\pgfkeysvalueof{/blitz/ratio}*\pgfkeysvalueof{/blitz/height}}\advance \pgf@y by \pgf@ya}
\anchor{south}{\centerpoint\pgfmathsetlength\pgf@ya{-(1-\pgfkeysvalueof{/blitz/ratio})*\pgfkeysvalueof{/blitz/height}}\advance \pgf@y by \pgf@ya}
  \anchor{east}{\centerpoint\pgfmathsetlength\pgf@ya{\pgfkeysvalueof{/blitz/breadth}}\advance \pgf@y by \pgf@ya\pgfmathsetmacro{\alpha}{atan(2*\pgfkeysvalueof{/blitz/ratio}*\pgfkeysvalueof{/blitz/height}/\pgfkeysvalueof{/blitz/width})}\pgfmathsetlength\pgf@xa{.5*\pgfkeysvalueof{/blitz/width}+\pgfkeysvalueof{/blitz/breadth}/tan(\alpha/2)}\advance\pgf@x by \pgf@xa}
  \anchor{west}{\centerpoint\pgfmathsetlength\pgf@ya{\pgfkeysvalueof{/blitz/breadth}}\advance \pgf@y by -\pgf@ya\pgfmathsetmacro{\alpha}{atan(2*\pgfkeysvalueof{/blitz/ratio}*\pgfkeysvalueof{/blitz/height}/\pgfkeysvalueof{/blitz/width})}\pgfmathsetlength\pgf@xa{.5*\pgfkeysvalueof{/blitz/width}+\pgfkeysvalueof{/blitz/breadth}/tan(\alpha/2)}\advance\pgf@x by -\pgf@xa}

  \backgroundpath{
    \centerpoint \pgf@xa=\pgf@x \pgf@ya=\pgf@y
    \pgfmathsetmacro{\alpha}{atan(2*\pgfkeysvalueof{/blitz/ratio}*\pgfkeysvalueof{/blitz/height}/\pgfkeysvalueof{/blitz/width})}
    \pgfmathsetlength\pgf@xb{.5*\pgfkeysvalueof{/blitz/width}-\pgfkeysvalueof{/blitz/breadth}/tan(\alpha/2)}
    \pgfmathsetlength\pgf@yb{\pgfkeysvalueof{/blitz/breadth}}
    \advance \pgf@xa by \pgf@xb
    \advance \pgf@ya by -\pgf@yb
\pgfpathmoveto{\pgfpoint{\pgf@xa}{\pgf@ya}}
    \pgfmathsetlength\pgf@xb{\pgfkeysvalueof{/blitz/width}}
    \advance \pgf@xa by -\pgf@xb
    \pgfpathlineto{\pgfpoint{\pgf@xa}{\pgf@ya}}
    \pgfmathsetlength\pgf@yb{\pgfkeysvalueof{/blitz/breadth}+\pgfkeysvalueof{/blitz/ratio}*\pgfkeysvalueof{/blitz/height}}
    \pgfmathsetlength\pgf@xb{\pgf@yb*sin(90-\alpha)}
    \advance \pgf@xa by \pgf@xb
    \advance \pgf@ya by \pgf@yb
    \pgfpathlineto{\pgfpoint{\pgf@xa}{\pgf@ya}}
    \pgfmathsetlength\pgf@xb{2*\pgfkeysvalueof{/blitz/breadth}/cos(90-\alpha)}
    \advance \pgf@xa by \pgf@xb
    \pgfpathlineto{\pgfpoint{\pgf@xa}{\pgf@ya}}
    \pgfmathsetlength\pgf@yb{\pgfkeysvalueof{/blitz/ratio}*\pgfkeysvalueof{/blitz/height}-\pgfkeysvalueof{/blitz/breadth}}
    \pgfmathsetlength\pgf@xb{\pgf@yb*sin(90-\alpha)}
    \advance \pgf@xa by -\pgf@xb
    \advance \pgf@ya by -\pgf@yb
    \pgfpathlineto{\pgfpoint{\pgf@xa}{\pgf@ya}}
    \pgfmathsetlength\pgf@xb{\pgfkeysvalueof{/blitz/width}}
    \advance \pgf@xa by \pgf@xb
    \pgfpathlineto{\pgfpoint{\pgf@xa}{\pgf@ya}}
    \pgfmathsetlength\pgf@yb{(1-\pgfkeysvalueof{/blitz/ratio})*\pgfkeysvalueof{/blitz/height}+\pgfkeysvalueof{/blitz/breadth}}
    \pgfmathsetlength\pgf@xb{.5*\pgfkeysvalueof{/blitz/width}+\pgfkeysvalueof{/blitz/breadth}/tan(\alpha/2)}
    \advance \pgf@xa by -\pgf@xb
    \advance \pgf@ya by -\pgf@yb
    \pgfpathlineto{\pgfpoint{\pgf@xa}{\pgf@ya}}
    \pgfpathclose
  }
}

\newcommand*{\srefset}{\pgfqkeys{/sref}}\srefset{
  height/.initial=.5cm,
  width/.initial=.5cm,
}

\pgfdeclareshape{sref}
{
  \savedanchor\centerpoint{
    \pgf@x = .5\wd\pgfnodeparttextbox
    \pgf@y = .5\ht\pgfnodeparttextbox
  }
  \anchor{center}{\centerpoint}
  \anchor{north}{\centerpoint\pgfmathsetlength\pgf@ya{.5*\pgfkeysvalueof{/sref/height}}\advance \pgf@y by \pgf@ya}
  \anchor{south}{\centerpoint\pgfmathsetlength\pgf@ya{-.5*\pgfkeysvalueof{/sref/height}}\advance \pgf@y by \pgf@ya}
  \anchor{east}{\centerpoint\pgfmathsetlength\pgf@xa{.5*\pgfkeysvalueof{/sref/width}}\advance \pgf@x by \pgf@xa}
  \anchor{west}{\centerpoint\pgfmathsetlength\pgf@xa{-.5*\pgfkeysvalueof{/sref/width}}\advance \pgf@x by \pgf@xa}

  \backgroundpath{
    \centerpoint \pgf@xa=\pgf@x \pgf@ya=\pgf@y
    \pgfmathsetlength\pgf@xb{.5*\pgfkeysvalueof{/sref/width}}
    \pgfmathsetlength\pgf@yb{.5*\pgfkeysvalueof{/sref/height}}
    \pgfmathsetlength\pgf@yc{.75*\pgfkeysvalueof{/sref/height}}
    \advance \pgf@xa by \pgf@xb
    \pgfpathmoveto{\pgfpointpolar{20}{\pgf@xa}}
    \pgfpatharc{20}{170}{\pgf@xb}
    \pgfpathlineto{\pgfpointpolar{140}{\pgf@yc}}
    \pgfpathmoveto{\pgfpointpolar{200}{\pgf@xa}}
    \pgfpatharc{200}{350}{\pgf@xb}
    \pgfpathlineto{\pgfpointpolar{320}{\pgf@yc}}
  }
}

\newcommand*{\qmarkset}{\pgfqkeys{/qmark}}\qmarkset{
  height/.initial=10pt,
  width/.initial=10pt,
  ratio/.initial=.5,
}

\pgfdeclareshape{qmark}{
  \savedanchor\centerpoint{\pgf@x=.5\wd\pgfnodeparttextbox \pgf@y=.5\ht\pgfnodeparttextbox }
  \anchor{center}{\centerpoint}
  \anchor{north}{\centerpoint\pgfmathsetlength\pgf@ya{\pgfkeysvalueof{/blitz/ratio}*\pgfkeysvalueof{/blitz/height}}\advance \pgf@y by \pgf@ya}
  \anchor{south}{\centerpoint\pgfmathsetlength\pgf@ya{-(1-\pgfkeysvalueof{/blitz/ratio})*\pgfkeysvalueof{/blitz/height}}\advance \pgf@y by \pgf@ya}
  \anchor{west}{\centerpoint\pgfmathsetlength\pgf@xa{-.5*\pgfkeysvalueof{/blitz/height}}\advance \pgf@x by \pgf@xa}
  \anchor{east}{\centerpoint\pgfmathsetlength\pgf@xa{.5*\pgfkeysvalueof{/blitz/height}}\advance \pgf@x by \pgf@xa}

  \backgroundpath{
    \pgfkeysgetvalue{/qmark/width}{\width}
    \pgfkeysgetvalue{/qmark/height}{\height}
    \pgfkeysgetvalue{/qmark/ratio}{\r}
    \pgfpathmoveto{\pgfpointadd{\centerpoint}{\pgfpoint{-.5*\width}{-.1*\height}}}
    \pgfpathcurveto
{\pgfpointadd{\centerpoint}{\pgfpoint{-3/6*\width}{.3*\height}}}
      {\pgfpointadd{\centerpoint}{\pgfpoint{.5*\width}{.4*\height}}}
      {\pgfpointadd{\centerpoint}{\pgfpoint{.5*\width}{-.05*\height}}}
    \pgfpathcurveto
{\pgfpointadd{\centerpoint}{\pgfpoint{.5*\width}{-.45*\height}}}
      {\pgfpointadd{\centerpoint}{\pgfpoint{-.17*\width}{-.2*\height}}}
      {\pgfpointadd{\centerpoint}{\pgfpoint{-.2*\width}{-0.78*\height}}}
    
    \pgfpathcircle{\pgfpointadd{\centerpoint}{\pgfpoint{-.2*\width}{-1.0*\height}}}{.08*\width}
  }
}

\makeatother

\makeatletter
\@ifclassloaded{beamer}
  {}
  {

\theoremstyle{remark}
    
    \theoremstyle{definition}

  }
\makeatother

\newif\ifradical
\radicaltrue
\newif\ifbeta
\betatrue
\newif\ifuncertain
\uncertaintrue
\newif\ifcomments
\commentstrue

\newcounter{CtrSprachspiel}

\newcommand{\mws}{\ensuremath{M_W^{S}}}
\newcommand{\mwf}{\ensuremath{M_W^{F}}}
\newcommand{\mfs}{\ensuremath{M_F^{S}}}
\newcommand{\mwfs}{\ensuremath{M_W^{S\!F}}}
\newcommand{\pProj}[1]{\ensuremath{\mathfrak{P}'(#1)}}
\newcommand{\pProjH}{\ensuremath{\pProj{\H}}}

\makeatletter\begin{document}

\title{Reflecting on the ``Measurement'' Problem}
\title{The Measurement Syndrome}
\title{Wigner's Isolated Friend}

\author{Arne Hansen and Stefan Wolf}
\affiliation{Facolt\`a di Informatica, 
Universit\`a della Svizzera italiana, Via G. Buffi 13, 6900 Lugano, Switzerland}

\date{\today}
  
\begin{abstract}
\noindent
  The measurement problem is seen as an ambiguity of quantum mechanics, or, beyond that, as a contradiction \emph{within} the theory:
Quantum mechanics offers two conflicting descriptions of the Wigner's-friend experiment.
As we argue in this note there are, however, obstacles from \emph{within} quantum mechanics and regarding our perspective onto \emph{doing physics} towards \emph{fully describing a measurement}.
We conclude that the ability to exhaustively describe a measurement is an \emph{assumption} necessary for the common framing of the measurement problem and ensuing suggested solutions. 
 \end{abstract}

\maketitle

\section{Introduction}\label{sec:introduction}
\noindent
If one attempts to include observers into the quantum-mechanical description, one encounters problems such as the measurement problem~\cite{Maudlin95,BHW16}.
The Wigner's-friend experiment~\cite{wigner1963problem, deutsch1985quantum, Wigner1961} is a manifestation thereof:
\emph{Wigner} measures his friend who, in turn, measures another system.
With appropriately chosen measurements, the friend ends up being in a superposition of states after his measurement.
This is at odds with the idea that measurements yield definite results.
The definiteness is commonly translated to the case of Wigner's friend by demanding that the joint system~$F\otimes S$ collapses to one of a number of orthogonal states representing different measurement results.
In order to obtain a formal contradiction, it is crucial to describe the friend's measurement by means of quantum mechanics.
This amounts to the following statement:
\begin{enumerate}[label={(A)}]
  \item\label{assum} For an observer~$O$ observing a result~$x$ when measuring a system~$S$, it is sufficient that~$O\times S$ is in a state~$\phi$ for some~$\phi\in\H_O\otimes\H_S$.
\end{enumerate}
The measurement problem in its usual reading results from the discrepancy between~$\phi_{\text{uni}}=U\phi_{\text{init}}$, i.e., the result of a unitary evolution~$U$, and~$\phi_{\text{cps}}$, i.e., the ``collapsed'' post-measurement state associated with ``definite measurement results'' if one chooses a suitable~$\phi_{\text{init}}$ and~$U$.

The association of ``having measured~$x$'' with ``being in a state~$\phi$'' is justified if physical theories are taken to describe an ontological reality.
We adopt the weaker position that physics provides us with descriptions gauged by our (experimentally acquired) experience.
Thus, we start from the following two characteristics which we regard as necessary~(albeit not sufficient) for \emph{doing physics}:
\begin{enumerate}[label={(C\arabic*)}]
  \item\label{c1} Physics strives for a (formal) \emph{description} of the world.\item\label{c2} \emph{Experience} provides the basis for normative judgements about the correctness of the theory.\end{enumerate}
Then, the association expressed in~\ref{assum} faces obstacles that we examine in this note:
How can a general, contextual theory establish reference to a particular system?
How can we expect to exhaustively describe the measurement while maintaining its role in deciding about the correctness of a theory?
In light of these questions, the statement~\ref{assum} constitutes an \emph{implicit assumption} necessary to seeing a problem in the measurement.

In Section~\ref{sec:setup}, we recapitulate the measurement problem.
In Section~\ref{sec:the_system}, we raise issues around the notion of a ``system,'' before we turn to the description of the friend's measurement in Section~\ref{sec:the_friend}.
Finally, in Section~\ref{sec:assumptions_and_contingencies}, we discuss the previously excavated assumption and its import.

 \section{The setup}\label{sec:setup}
\noindent
When we refer to the Wigner's-friend experiment, we think of the following scenario:
Wigner ensures that the friend is ready to perform a measurement by means of a measurement~$\mwf$.\footnote{Subsequently, we adopt the following notation: The symbol~$M_{O}^{S}$ denotes a measurement performed by the observer~$O$ on the system~$S$.
Note that we refer to an observer~$O$ by a capital letter: In this regard,~$O$ is just another system and, thus, can be measured, too.
If we refer to measurements on joint systems~$S\otimes P$, we abbreviate~$M^{S\otimes P}_{O}$ by~$M^{S\!P}_{O}$.}
He then prepares a system~$S$ with an initial measurement~$\mws$ and sends it to his friend.
The friend performs a measurement~$\mfs$ on~$S$.
The joint system~$S\otimes F$ remains however isolated, and, therefore, evolves unitarily.
Finally, Wigner performs the measurement~$\mwfs$ on the joint system.
\begin{equation*}
  \begin{tikzpicture}
    \def\w{7cm}
\def\h{.9cm}
\draw[thick] (0,0) to coordinate[pos=.2] (mw11) coordinate[pos=1/2] (mw12) coordinate[pos=.8] (mw13) ++(\w,0);
\draw[thick] (0,-\h) to coordinate[pos=.2] (mw21) coordinate[pos=1/2] (mw22) coordinate[pos=.8] (mw23) ++(\w,0);
\draw[<-,shorten <=2pt] (0,0) to[out=180,in=-90] ++(-.5,.3) node[anchor=south] {$S$};
\draw[<-,shorten <=2pt] (0,-\h) to[out=180,in=90] ++(-.5,-.3) node[anchor=north] {$F$};
\draw[thick] (mw12) to (mw22);
\fill[] (mw22) circle(2pt);
\node[fill=orange!30, draw=orange, thick] (mws) at (mw11) {$M_W^{S}$};
\node[fill=orange!30, draw=orange, thick] (mfs) at (mw12) {$M_F^{S}$};
\node[fill=orange!30, draw=orange, thick] (mwf) at (mw21) {$M_W^{F}$};
\draw[fill=orange!30, draw=orange, thick] ({$(mw13)+(-.6,0)$} |- {mws.north}) coordinate (r1) rectangle ({$(mw23)+(.6,0)$} |- {mwf.south}) coordinate (r2);
\node[] at ($.5*(r1)+.5*(r2)$) {$M_W^{S\!F}$};
\draw[dashed,rounded corners] ($(mfs.north east)+(.2,.2)$) coordinate (r21) rectangle ({$(mwf.south)+(0,-.2)$} -| {$(mfs.west)+(-.2,0)$}) coordinate (r22);
\node[font=\footnotesize] at ($(mfs)+(0,-1.7)$) {unitary evolution/isolated system};
   \end{tikzpicture}
\end{equation*}

Commonly, the measurement~$\mws$ is replaced by a phrase of the sort ``Wigner prepares a system in a state~$\phi\in\H$.''
We grant Wigner the ability to choose whether to send the system~$S$ to the friend or not conditioned on the result of his measurement.
Further, we allow for measurements such that eigenvalues to projectors orthogonal to~$\proj{\phi}$ are different from the eigenvalue~$\lambda$ associated with~$\phi$.
Thus, when ``Wigner prepares~$S$ in a state~$\phi$,'' he merely sends~$S$ to his friend if the measurement~$\mws$ yields the result~$\lambda$.
Similarly, the statement ``The friend is ready to perform his measurement.'' means that Wigner knows that ``the friend is ready.''
If Wigner aims to describe his friend by means of quantum mechanics, then he gets to know that ``the friend is ready'' by performing a measurement~$\mwf$.

The measurement problem in this setup amounts to the following: 
Quantum mechanics ostensibly allows for two different descriptions of the friend's measurement.
For suitably chosen measurements, these result in different predictions for the probability distribution of Wigner's final measurement~$\mwfs$.
If we ascribe one of these descriptions to the friend, and the other to Wigner, then we conclude that they disagree on the predictions for Wigner's final measurement.
Straight solutions\footnote{A \emph{straight} solutions consists of \textcquote[p.~69]{KripkeWRPL}[]{pointing out to the silly sceptic a hidden fact he overlooked}, i.e., that (at least) one of the assumptions is unwarranted.}
to the measurement problem consist of restricting quantum mechanics as to not apply to measuring observers, to preclude the system~$F$ to be isolated, to choose one of the available descriptions as the correct one, or to embrace the predictions of quantum mechanics as subjective.

In the following, we examine what it takes to ``measure that the friend is ready,'' and what it takes to describe his ensuing measurement.
We consider first challenges that arise within quantum mechanics and then issues that affect \emph{any} physical theory.
 \section{The system}\label{sec:the_system}
\noindent
The system~$S$ exposes issues with the \emph{physical systems in general}.
Wigner and his friend must agree to refer to the \emph{same} system~$S$.
In the above scenario, this amounts to Wigner reading off by means of his measurement~$\mwf$ that the friend refers to the same~$S$ as he does.
If the friend is ready to ``measure~$S$,'' then the friend ``has~$S$ in mind.''
We have to add to~\ref{assum} the following aspect:
\begin{enumerate}[label={(ASys)}]
\item\label{AssumSys} For an observer~$O$ referring to a system~$S$, it is sufficient that~$O$ is in a state~$\phi$ for some~$\phi\in\H_O$.
\end{enumerate}

Before examining whether a measurement can establish reference to a system, let us take a closer look at the notion of a ``system'' itself.

\emph{The notion builds on the idea that we can talk about a clearly confined part of the world around us, and distinguish it from other such parts.}
We assume a \emph{separability}, i.e., the possibility to make statements about one such part independent of other parts.
Einstein places this separability at the core of his understanding of \emph{physical reality}:\footnote{We omit the following part of the quote, which, in light of non-locality~\cite{Bell1964,aspect1982experimental,ma2012experimental} is questionable. 
\blockcquote[\S5 (translated quote from~\cite{EinsteinBorn})]{sep-einstein-philo}[.]{If one adheres to this program, then one can hardly view the quantum-theoretical description as a complete representation of the physically real. If one attempts, nevertheless, so to view it, then one must assume that the physically real in~$B$ undergoes a sudden change because of a measurement in~$A$. My physical instincts bristle at that suggestion}
We do not regard separability as \emph{sufficient} in the sense that measurements on parts, together with previously shared information, reveal the results of any measurement that can possibly be performed on the combined system.}
\blockcquote[\S5 (translated quote from~\cite{EinsteinBorn})]{sep-einstein-philo}[.]{I just want to explain what I mean when I say that we should try to hold on to physical reality. 
We are, to be sure, all of us aware of the situation regarding what will turn out to be the basic foundational concepts in physics: the point-mass or the particle is surely not among them; the field, in the Faraday-Maxwell sense, might be, but not with certainty. 
But that which we conceive as existing (`actual') should somehow be localized in time and space. That is, the real in one part of space,~$A$, should (in theory) somehow `exist' independently of that which is thought of as real in another part of space,~$B$. If a physical system stretches over the parts of space~$A$ and~$B$, then what is present in~$B$ should somehow have an existence independent of what is present in~$A$. What is actually present in~$B$ should thus not depend upon the type of measurement carried out in the part of space,~$A$; it should also be independent of whether or not, after all, a measurement is made in~$A$.
[\ldots]
However, if one renounces the assumption that what is present in different parts of space has an independent, real existence, then I do not at all see what physics is supposed to describe. For what is thought to be a `system' is, after all, just conventional, and I do not see how one is supposed to divide up the world objectively so that one can make statements about the parts}
Importantly, the ability to speak of separate systems does not imply an ``objective division'' of the world.
Einstein's reservations towards such a division go beyond a temporal inability to conceive such an objective division:
\blockcquote[p.~102, own translation]{Einstein1916}[.]{Terms that have proven useful for the ordering of things attain easily such an authority over us so that we forget their worldly origin and we accept them as unalterable facts.
They are, then, put down as `thinking-necessities,' `a priori given,' etc.
The path of scientific progress is often made impassable for a long time by such misconceptions}
 This joins Feyerabend's~\cite{FeyerWDMZ,FeyerZV} and Kuhn's~\cite{kuhn1962structure} investigation into the history of science, refuting the idea of an overarching convergent trend, with the considerations by Wittgenstein, Sellars, and Rorty~\cite{WittgPhiloUntersuchungen,Sellars1956,RortyCIS} on the contingency of language.
If we assume that there is neither a final privileged language, i.e., a \textcquote{RortyCIS}[]{truth out there}, nor that we are able to \textcquote{RortyCIS}[]{step outside the various vocabularies we have employed}, then we must allow for a Kuhnian paradigm shift, i.e., a radical re-description.
Adopting~\ref{assum} and~\ref{AssumSys} is, in turn, a step towards assuming a privileged language, at the risk of petrifying scientific discourse.
But what supports this suspicion towards the existence of, or convergence towards, an ultimate language that reflects the truth out there other than assuming uniformity in the history of science?\footnote{\label{fn:hist_arg}Arguments like the following-rule paradox against the existence of a privileged language (see Section~\ref{sub:privat_language_monologistic_science}) are based on the observation that such uniformity is not warranted. 
Thus, historical arguments are tainted.}
What suggests to repudiate~\ref{AssumSys}?

\subsection{Isolated systems, decoherence, and superselection rules}\label{sub:isolated_systems}
\noindent
A problem with reference to systems is related to decoherence in the following sense: 
If Wigner observers a non-unitary time evolution of~$S\otimes F$, then it is possible that the friend measures a system~$S'$ bigger than~$S$, i.e., bigger than what Wigner thinks the system is.
The ``escaping photon'' leading to decoherence can be seen as a problem of non-aligned reference:
The photon is contained in~$S'$ but not in~$S$.
Conversely, we can see decoherence---provided that we take it as \emph{inevitable} interaction with the environment---as the inability to sharply draw the boundary between one system and another, or between one system and its environment.
This perspective also challenges how we understand \emph{decoherence} itself: 
We cannot define decoherence as ``a system interacting with its environment'' because there is no clear-cut distinction between ``system'' and ``environment.''
\emph{Thus, decoherence is maybe better seen as the abandonment of the notion of a well-confined system.}

This also affects \emph{environment-induced superselection rules}.
The program of ``einselection'' addresses the problem of many-worlds interpretations how to fix the basis corresponding to measurement results, and, thus, how ``to split the worlds.''
Einselection is not primarily concerned with explaining how a single world splits into systems.
Explaining the system-split yet poses a problem.
Superselection rules might provide a basis on which one might attempt to divide the world into systems, against Einstein's above-mentioned concerns.
These superselection rules cannot be induced by the environment,\footnote{On a side-note, let us remark that the idea of ``being induced by the environment'' resembles conceptually the epistemic idea of sense-data: There is something \emph{given} in our environment that induces us to know (see later footnotes.)} because it assumes the notion of a system already, leaving us with a circularity. We are similarly faced by circularity, if we attempted to ``measure'' what qualifies as a system, because ``measuring'' here means again ``measuring a system.''
A way out is to supplement quantum mechanics from the beginning by superselection rules.

The idea of introducing superselection rules has another interesting consequence:
If we regard contextuality as an essential aspect of quantum mechanics~\cite{Piron1976,Gudder1968,KS67}, then the superselection rules effectively undermine quantum mechanics itself: 
With the superselection rules, we introduce observables in the center of the orthomodular lattice of allowed projectors~$\pProjH$, i.e., elements that commute with all other elements in the lattice.
This is \emph{the cost of introducing a non-contextual notion of a system into contextual theory}~\cite{Piron1976}:
There must be a \emph{Heisenberg cut}, i.e., a line at which things become at least in part classical.
This might also affect the measurement problem directly:
If in the Wigner's-friend experiment the notion of a system was established by suitable superselection rules, not all of the measurements leading to the contradiction are necessarily permitted.
If, for instance, the measurement~$\mwfs$ with ambiguous probability predictions was precluded by the superselection rules, then there would be no measurement problem.\footnote{This reminds of the Bohmian restriction to position measurements.}

There are objections: Quantum mechanics with postulated superselection rules is \emph{not} quantum mechanics.
Or, put otherwise, introducing superselection rules undermines the assumption that quantum mechanics is \emph{universal}.\footnote{Consider the following extreme case:
Let us assume the world is entirely classical. 
Then, we might still describe it by a Boolean sub-lattice of $\ProjH$.
Is then quantum mechanics universal, given that we have to restrict the quantum mechanical description to a Boolean sub-lattice?}
We are left with a problem: How can quantum mechanics be universal and exhaustive without threatening the notion of a system?
How can quantum mechanics \emph{be} without the notion of system?
How can the notion of a system avoid the above circularities without supplementing the theory by such a notion and threatening its universality?
This necessarily levels all particular features of a particular system.\footnote{See also the discussion in Section~\ref{sub:privat_language_monologistic_science}.}
 \section{The Friend}\label{sec:the_friend}
\subsection{Quantum inquiries about intentions}\label{sub:quantum_inquiries_about_intention}
\noindent
Let us assume that, despite the above scepticism, there are theoretical means to \emph{define} a system, e.g., by a suitable set of superselection rules added to quantum mechanics.
The question remains whether there can exist an element~$\Pi_W^{F}\in\Proj{\H_{F}}$ that shows that ``the friend means to measure~$S$''---that the friend ``has~$S$ in mind.''
This question carries two intricacies: Can we find something \emph{inside} the friend that reveals
\begin{enumerate}[label={(Itnt\arabic*)}]
  \item\label{Intention1} the friend's reference to something outside of him, and
  \item\label{Intention2} the friend's intention to perform a measurement?
\end{enumerate}
If we require that the friend's intention to measure~$S$ is before his actual contact with~$S$, then we must expect to read off this intention by merely measuring~$F$, and \emph{not}~$S\otimes F$. 

We examine the possibility of asking the more specific question whether 
\begin{enumerate}[label={(Itnt3)}]
  \item\label{Intention3} the friend intends to perform the measurement~$\Pi_F^S\in\Proj{\H_S}$.
\end{enumerate}
If Wigner can inquire about such a question, and answer it positively, then he can conclude that the friend has the intention ``to measure~$S$.''
It might be that one measurement~$\Pi_W^{F}$ with a suitable result reveals that the friend intends to ``perform the measurement~$\Pi$ on~$S$ for some~$\Pi\in P\subset\Proj{\H_S}$.''
Let us, therefore, assume the map 
\begin{equation*}
  \pi : \Proj{\H_S} \to \Proj{\H_F}
\end{equation*}
to formally capture the following: 
If Wigner can positively answer to~$\Pi_W^F\in\Proj{\H_W}$, then he can conclude that the friend intends to measure an element in~$\pi^{-1}(\Pi_W^F)$.

The friend does not intend to measure \emph{because} we asked him about it.
Or, even worse, oscillate between different intentions, as we change our inquiries.\footnote{\label{fn:ice_cream}
  In a sense, this is what we insinuated when we allowed for concluding from~\ref{Intention3} to~\ref{Intention1} and~\ref{Intention2}.
  To illustrate this, consider the following situation:
  You offer your friend one scoop of ice-cream, and you tell him to choose one flavor.
  To find out which flavor he would like to have, you ask him the following two questions (arbitrarily many times and in arbitrary order):
  \begin{enumerate*}[label={(Q\arabic*)}]
    \item\label{qchoc} ``Do you want chocolate ice-cream?'' and 
    \item\label{qstrb} ``Do you want vanilla ice-cream?''
  \end{enumerate*}
  If he now says consistently ``yes'' to one, and ``no'' to the other, then you might conclude that he has a clear ``intention.''
  This conclusion is not justified if your friend gives changing answers as you repeat these questions.
  If he says consistently ``yes'' to both questions, then you doubt that he understood what you meant with ``choosing one flavor.''
  If he does not understand what it means to ``choose a flavour,'' how can we conclude his intentions from his answers to the questions~\ref{qchoc} and~\ref{qstrb}?

  Even though we are getting a little ahead of ourselves here, let us take this one step further and imagine that you give him a cone with a scoop of ice-cream. 
  Your friend starts eating or not.
  In the case that the friend consistently gives a preference, and you offer him the respective ice-cream, you expect him to starting eating (and his face showing how he is enjoying it).
  Conversely, if you offer him the other flavor, then he should not eat and rather insist ``But I meant chocolate ice-cream.''
  Now, if you offer him the type that you concluded from his consistent answers to be his preference, and he does not start eating, we can imagine the friend to say ``Oh sorry, I meant the vanilla.'' or, angrily ``But I meant chocolate.''
  In the latter case you might learn that he (consistently) permutes the words vanilla and chocolate.
  If your friend in the past always asked for vanilla, and always ate it with joy, and you now skip the question to directly offer him vanilla ice-cream, then he might still not eat it, e.g., because today he ``felt like chocolate,'' and, thus, intended to have chocolate ice-cream.
  In the case of the friend giving changing answers, we can imagine that he (emotionlessly) eats the ice-cream, unless he said ``no'' and you give him the respective flavor nonetheless.
  Does this last case, despite being logically correct, still qualify for ``having chosen a flavor''?
  Now, one might oppose that I deprived the friend of his emotions, to testify of his choice.
  But then I can easily imagine him showing disgust, even if he eats the logically correct flavor, e.g., because \emph{this} chocolate ice-cream is horrible.
  Did I now fiddle with the \emph{circumstances} too much? Should there not be a clear pattern, if we consider ``normalized conditions''?
  But how are these conditions characterized? 
  By the friend sticking to the logical rule? 
  This leaves us with a circularity.
  These are first steps towards the discussion in Section~\ref{sub:privat_language_monologistic_science}.

}
We, therefore, demand that the image~$\pi(\Proj{\H_s})$ forms a distributive lattice avoiding that the friend ``changes his mind'' if we ask him questions in a different order. 
Thus,~$\pi$ effectively collapses the non-distributive lattice~$\Proj{\H_S}$ into a distributive sub-lattice of~$\Proj{\H_F}$.
Thus, we re-encounter the preferred basis problem of the many-worlds interpretation and the ``system'' problem from Section~\ref{sub:isolated_systems}:
We are required to postulate a distributive sub-lattice that corresponds to ``measuring intentions.''
Suspecting that the friend has ``$S$ in (his) mind'' posits intentions as non-contextual.
As in Section~\ref{sub:isolated_systems}, we encounter issues anchoring non-contextual notions in a contextual theory that is supposedly universally valid.
Thus, we do not have to go as far as the measurement problem for contextuality\footnote{This takes contextuality as essential to the measurement problem (see~\cite{HW19GMP}).} to get into the way of an exhaustive quantum description.

\subsection{Normativity and the nexus\\ between past and future}\label{sub:privat_language_monologistic_science}
\noindent
The above-developed scepticism relates more specifically to quantum mechanics.
The requirements~\ref{c1} and~\ref{c2} give rise to another, more general doubt.
Recall that~\ref{c1} specifies the \emph{descriptive task} of physics, while~\ref{c2} specifies that this description has to \emph{succumb to a normative judgement}.
Kripke, when discussing Wittgenstein's paradox of rule-following\footnote{Wittgenstein summarizes:
\blockcquote[\S201]{WittgPhiloInvestigations}[.]{This was our paradox: no course of action could be determined by a rule, because every course of action can be made out to accord with the rule.
The answer was: if everything can be made out to accord with the rule, then it can also be made out to conflict with it.
And so there would be neither accord nor conflict here}
 To put it in the words of the example in Footnote~\ref{fn:ice_cream}, imagine that your friend in the past always asked for vanilla ice-cream. 
We cannot tell whether he is following a rule that dictates him to choose vanilla again, or a rule that lets him change his choice for chocolate ice-cream this time, \emph{before} he has uttered his choice.
}, observes:
\blockcquote[p.37, emphasis in original]{KripkeWRPL}[.]{The relation of meaning and intention to future action is \emph{normative}, not \emph{descriptive}}
On the one hand, this statement reflects on the above-discussed reference to systems and the problem of Wigner aligning his reference to his friend's.
On the other hand, it foreshadows a similar problem with \emph{describing} the friends measurement---i.e., referring to the same measurement that supposedly provides the normative experience. 
If we \textcquote{Mueller17}[]{interchangeably use the words `experience', `observation,' and `state of the observer'\,}, then we remove the room for normative judgement. 
We effectively postulate the theory to be true.\footnote{This brings us back to reflect Einstein's concerns about the passibility on the path of scientific progress.}
In the following, we elaborate on these concerns.

There is a link between Hume's problem of induction and Wittgenstein's paradox of rule-following:
\blockcquote[p.~62, emphasis in original]{KripkeWRPL}[.]{Both [Hume and Wittgenstein] develop a sceptical paradox, based on questioning a certain \emph{nexus} from past to future.
Wittgenstein questions the nexus between past `intention' or `meanings' and present practice: for example, between my past `intentions' with regard to `plus' and my present computation `68+57=125'.
Hume questions two other nexuses, related to each other: the causal nexus whereby a past event necessitates a future one, and the inductive inferential nexus from the past to the future}
 The link between the two can be explicated as: \emph{No\-thing in the friend's past logically determines whether and how the friend means to account for his experience in a measurement.}
That is:
Nothing in the friend's past determines whether he intends to measure~$S$ or~$T$, or whether he thinks to have successfully performed a measurement, or what result he obtained.
Unless the friend's particular state~$a$ can be subsumed under a general category~$A$.
The extension into the future of this \emph{general} category is not logically warranted~\cite{sep-induction-problem}.
Thus, all theories have a tentative character and we must resort to falsification~\cite{Popper1934}.

The idea that quantum mechanics provides means to exhaustively describe the friend's measurement is to assume that quantum mechanics provides the general category that covers all future measurements.
Then, however, the friend is no different from a brain-in-a-vat~\cite{putnam1981}: 
Employing the separability assumption from above, we can conclude that anything that we can say about the friend, including his intentions and possible accounts of experience, can be derived from the friend's state.
With the help of some auxiliary environment~$T$, we can simulate the friend's measurement without ever putting him in contact with~$S$ despite his alleged initial intentions of measuring~$S$.\footnote{We merely have to ensure the partial trace on~$\H_F$ to remain the same.}
This renders the reference to the~$S$ that we struggled to ensure previously---as a necessary requirement for qualifying the friends acts as a measurement of~$S$---invalid.
The friend's reference does not conform to our notion of ``referring to~$S$,'' as in referring to \emph{a particular} system~$S$.
There does not seem to be a general criterion for particular reference~\cite{putnam1991representation}.
The problem translates to any other observer, also to us:\footnote{If quantum mechanics fully describes the friend's measurement, what then is ``reading this text'' other than a quantum-mechanically described observation?}
If there is no principle difference between the friend and us, we are thrown back to the question how we can ever refer to anything outside ourselves, if we must suspect some Wigner-like super-observer simulating us.
Similarly, a measurement is no more something we actually, or better \emph{actingly}, have a part in, but something that \emph{is said about us}.\footnote{In this regard, ``measuring'' becomes ``being said to measure,'' analogue to~\cite[\$202]{WttgstnOgden90}.}
Is there then any measurement that could provide us with the grounds to reject a theory as demanded in~\ref{c2}?

Kripke's argument against functionalism can be seen as a variant of the following-rule paradox.
Translating it to Wigner's friend allows to explicate the issue.
Let us first summarize the argument laid out in~\cite{Buechner2018}:
If we assume that physical computers can \emph{break down}, then we can also imagine the following scenario: 
If a physical computer computes~$F$ and breaks down, it actually computes \emph{another} function~$G$.
If, vice versa, a computer computes~$G$ and breaks down, it actually computes the function~$F$.
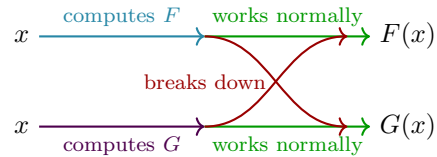
\begin{equation*}
  \begin{tikzpicture}
    \def\h{2.2}
\def\v{1.2}
\draw[thick, cyan!60!black, ->] (0,0) node[anchor=east,black] {$\strut x$} to node[midway, above, font=\footnotesize] {computes $F$} ++(\h,0) coordinate (mw1);
\draw[thick, green!60!black, ->] (mw1) to node[midway, above, font=\footnotesize] {works normally} ++(\h,0) coordinate (ep1) node[anchor=west,black] {$\strut F(x)$};
\draw[thick, violet!60!black, ->] (0,-\v)node[anchor=east,black] {$\strut x$} to node[midway, below, font=\footnotesize] {computes $G$} ++(\h,0) coordinate (mw2);
\draw[thick, green!60!black, ->] (mw2) to node[midway, below, font=\footnotesize] {works normally} ++(\h,0) coordinate (ep2)node[anchor=west,black] {$\strut G(x)$};
\draw[thick, red!60!black, ->] (mw1) to[out=0,in=180] ($(ep2)+(-.3,0)$);
\draw[thick, red!60!black, ->] (mw2) to[out=0,in=180] node[left,font=\footnotesize] {breaks down} ($(ep1)+(-.3,0)$);
   \end{tikzpicture}
\end{equation*}
Thus, we are left with the following problem:
\begin{quote}
  We cannot decide whether a physical computer physically computes~$\left\{\begin{array}{@{}c@{}}F\\ G\end{array}\right\}$, and~$\left\{\begin{tabular}{@{}l@{}}works normally\\ breaks down\end{tabular}\right\}$.
\end{quote}
We can stipulate that the computer works fine, and conclude that it computes~$F$ or~$G$.
Or we stipulate that it computes~$F$, and conclude whether it works correctly or breaks down.
But we lack the means to fix both.

The problem translates to the friend and his reference to systems~$S$ and~$T$:
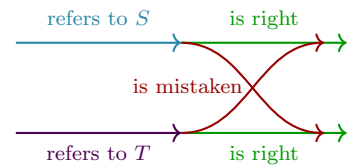
\begin{equation*}
  \begin{tikzpicture}
    \def\funcOne{refers to~$S$}
    \def\funcTwo{refers to~$T$}
    \def\worksNormally{is right}
    \def\breaksDown{is mistaken}
    \def\h{2.2}
\def\v{1.2}
\draw[thick, cyan!60!black, ->] (0,0) to node[midway, above, font=\footnotesize] {\strut\funcOne} ++(\h,0) coordinate (mw1);
\draw[thick, green!60!black, ->] (mw1) to node[midway, above, font=\footnotesize] {\strut\worksNormally} ++(\h,0) coordinate (ep1);
\draw[thick, violet!60!black, ->] (0,-\v) to node[midway, below, font=\footnotesize] {\strut\funcTwo} ++(\h,0) coordinate (mw2);
\draw[thick, green!60!black, ->] (mw2) to node[midway, below, font=\footnotesize] {\strut\worksNormally} ++(\h,0) coordinate (ep2);
\draw[thick, red!60!black, ->] (mw1) to[out=0,in=180] ($(ep2)+(-.3,0)$);
\draw[thick, red!60!black, ->] (mw2) to[out=0,in=180] node[left,font=\footnotesize] {\strut\breaksDown} ($(ep1)+(-.3,0)$);

   \end{tikzpicture}
\end{equation*}
If the friend is mistaken, then he refers to the respective other system.
\begin{quote}
  We cannot decide whether the friend refers to~$\left\{\begin{array}{@{}c@{}}S\\ T\end{array}\right\}$, and is~$\left\{\begin{tabular}{@{}l@{}}right\\ mistaken\end{tabular}\right\}$.
\end{quote}
So as long as we allow the friend to be mistaken at times---or dreaming, or hallucinating---, and this results in swapping references, or measurements,
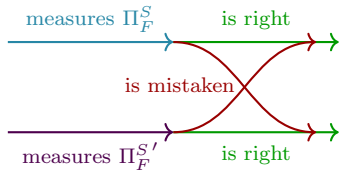
\begin{equation*}
  \begin{tikzpicture}
    \def\funcOne{measures~$\Pi^S_F$}
    \def\funcTwo{measures~${\Pi^S_F}'$}
    \def\worksNormally{is right}
    \def\breaksDown{is mistaken}
    \def\h{2.2}
\def\v{1.2}
\draw[thick, cyan!60!black, ->] (0,0) to node[midway, above, font=\footnotesize] {\strut\funcOne} ++(\h,0) coordinate (mw1);
\draw[thick, green!60!black, ->] (mw1) to node[midway, above, font=\footnotesize] {\strut\worksNormally} ++(\h,0) coordinate (ep1);
\draw[thick, violet!60!black, ->] (0,-\v) to node[midway, below, font=\footnotesize] {\strut\funcTwo} ++(\h,0) coordinate (mw2);
\draw[thick, green!60!black, ->] (mw2) to node[midway, below, font=\footnotesize] {\strut\worksNormally} ++(\h,0) coordinate (ep2);
\draw[thick, red!60!black, ->] (mw1) to[out=0,in=180] ($(ep2)+(-.3,0)$);
\draw[thick, red!60!black, ->] (mw2) to[out=0,in=180] node[left,font=\footnotesize] {\strut\breaksDown} ($(ep1)+(-.3,0)$);

   \end{tikzpicture}
\end{equation*}
then we are faced with the problem of deciding what measurement or what system the friend meant, \emph{and} whether he is right or mistaken.
There is nothing we can expect to find in the state of the friend that will definitely decide both question at once.
Again, we can stipulate that he is dreaming and establish whether he is referring to~$S$ or to~$T$.
Or we stipulate that he is referring to~$S$ and establish whether he is right or mistaken.
The authority to establish either of these stipulations is usually delegated to a wider circle of persons that \emph{intersubjectively} forms an agreement.\footnote{
  We can draw a connection to Habermas' criticism of positivism~\cite{HabermasEuI}:
  If quantum mechanics is taken to yield an exhaustive description, then the need for intersubjective agreement on norms is superfluous. 
  Language and science becomes private in a way that Habermas regards intrinsic to Pierce's pragmatism. 
  \blockcquote[{\S}6, p.~137, emphasis in original]{HabermasEuI_en}[.]{Peirce would have had to come upon the fact that the ground of \emph{intersubjectivity} in which investigators are always already situated when they attempt to bring about consensus about metatheoretical problems is not the ground of purpose-rational action, which is in principle solitary. [\ldots]
It is possible to think in syllogisms, but not to conduct a dialogue in them. [\ldots]
But the communication of investigators requires the use of language that is not confined to the limits of technical control over objectified natural processes.
It arises from symbolic interaction between societal subjects who reciprocally know and recognize each other as unmistakable individuals.
This communicative action is a system of reference that cannot be reduced to the framework of instrumental action}
 With the abandonment of intersubjectivity, Pierce cannot resort to a community of scientist vouching for methodologic means to yield certain knowledge as he effectively does: 
  \blockcquote[p.~109f, emphasis in in original]{HabermasEuI_en}[.]{Ontological propositions about the structure of reality unintentionally elucidate the process of mediation through which we come to know reality.
Yet in fact this concept of reality was first introduced only as the correlate of a process of inquiry that guarantees the cumulative acquisition of definitively valid statements.
As soon as we remember this point of departure, Scholastic realism of Peirce's stamp can be seen through as the ontologizing of an originally \emph{methodological} problem.
Indeed for Peirce the problem of the relation of the universal and the particular presented itself outside of the tradition.
That is, it appeared not as a logical-ontological problem, but rather in connection with the methodological concept of truth as a problem of the logic of inquiry}
   Irrespective of whether one adheres to Habermas' critical assessment of positivism or not, the connection illustrates the epistemological import of assuming an exhaustive language (see also Footnote~\ref{fn:sellars}).
}

In light of the challenges laid out above, we conclude:
The common reading of the measurement problem relies on \emph{fully describing} the friend's measurement, thus removing the friend from the ground of intersubjectivity.\footnote{This shows in an unquestioned use of phrases like the following (see Section~\ref{sec:setup}):
``If Wigner aims to describe his friend by means of quantum mechanics, then he gets to know that `the friend is ready' by performing a measurement~$\mwf$.''}

The ``problem'' then consists in the incommensurability of two different descriptions that quantum mechanics allows for.
\emph{The insistence that the measurement is a problem that needs a solution in form of another theory or an appropriate reading of quantum mechanics implicitly depends on removing the need to (intersubjectively) settle normative questions.}

What does this mean for the Wigner's-friend experiment?
No matter the reading of quantum mechanics,\footnote{We consider GRW~\cite{GRW} to be a different theory~\cite{BW17}.} an actual Wigner's-friend experiment surprises if an \emph{isolated system shows a collapse}---irrespective of what or who constitutes it:
The surprise would be that we can meaningfully call a system isolated \emph{despite} it showing a non-unitary evolution.
Whatever evidence justifies us to qualify the system as isolated is then at odds with quantum mechanics.
This takes $S\otimes F$ as just another quantum system.
Quantum mechanics is universal insofar as it can be applied to $S\otimes F$.
The ``isolated friend'' does, however, not provide any contribution to intersubjectively agreeing on the correctness of the theory.
Whatever happens inside $S\otimes F$ is \emph{not} a measurement that provides the experience required in~\ref{c2}.

Hume's problem of induction shows that one can hardly rely on uniformity to support general claims.
As such, the historical analysis of Feyerabend and Kuhn cannot serve as reason to reject the assumption of an exhaustive language as this requires a uniformity of history (see also Section~\ref{sec:the_system}, in particular Footnote~\ref{fn:hist_arg}).
The idea that science aims for its own disintegration strikes us, however, as odd:
For, can we not really learn when we have to listen carefully, or watch closely?
When mere description gives way to metaphorical disruption---\textcquote[\S1]{RortyCIS}[?]{suddenly breaking off the conversation long enough to make a face, or pulling a photograph out of your pocket and displaying it, or pointing at a feature of the surroundings, or slapping your interlocutor’s face, or kissing him} When experimental behavior is \emph{not} covered by the established description?
When emitted electrons do \emph{not} get any faster if the intensity of light shun onto a metallic plate is increased?

 \section{Assumptions and contingencies}\label{sec:assumptions_and_contingencies}
\noindent
If one accepts that the nexus from the past to the future poses an irreducible challenge (as discussed in Section~\ref{sec:the_friend}), then one accepts that the need for finding normative judgements persists.
The contingency of our description of the world out there is not a contemporary defect reflecting the inadequacy of our current description.
In this perspective, the characteristic~\ref{c2} remains an aspect of science.\footnote{With upholding~\ref{c2}, we repudiate the empiricist reading: ``\emph{Experience} provides the basis for judgements about the correctness of the theory.'' We do not believe that, once we figure out the ``language of sense-data,'' the room for normative judgements closes~(see, e.g.,~\cite{Sellars1956}).}
Science retains its room for creativity, and remains itself a creative activity~\cite{Fleck}.
These observations carry a circularity as they take us back to our starting point, i.e., the characteristic~\ref{c2} that we ascribed to physics in the first place.
Our arguments are subject to the contingency of language as well.
Thus, the problems we see in seeing the measurement as a problem \emph{are created} as much as, in our regards, is \emph{any} language, at least in part.
It seems that we must, at this point, retract from declaring our assumptions as ``weaker,'' as we initially did.
Such a comparison seems hardly justified.

If, on the contrary, one sees the quest of physics to excavate a \emph{truth out there}, independent of any normative judgements or creative acts---a truth that imposes itself---, then the assumptions~\ref{assum} and~\ref{AssumSys} can hardly be accepted as such, i.e., \emph{as assumptions}.
Identifying an assumption is a reflection on the creative steps that facilitate one's way of speaking:
It constitutes the admission that it could have been done otherwise.\footnote{\label{fn:sellars}There is the possibility to regard assumptions as a temporary evil until the evident foundations have been properly sorted out.
  Until the assumption is turned into an inevitable conclusion of self-imposing facts.
  This approach is, however, tainted by the tension within the idea of ``learning the self-imposing,'' or, to put otherwise, ``to acquire an unaquired ability.'' 
  This is the core of Sellars inconsistent triad: \blockcquote[\S6]{Sellars1956}[.]{[classical sense-datum theories] are confronted by an inconsistent triad made up of the following three propositions: 
A. $x$ senses red sense content $s$ entails $x$ non-inferentially knows that $s$ is red. 
B. The ability to sense sense contents is unacquired.
C. The ability to know facts of the form $x$ is {\o} is acquired}
 }
In this light, it is only consequent that positivism removes such reflections~(see~\cite{HabermasEuI}). 
We return to Wittgenstein's observation that the reader might not be easily convinced if he had not had similar thoughts before.
The ladder can merely be climbed if one has already made the first steps onto it~(see~\cite[preface and \S6.53]{witt:trac22}).
Embracing the contingencies of any language-game bars us from offering an ultimate argument for the contingency of language.\footnote{Putnam hints at this concern when he states: \blockcquote[\S5]{putnam1991representation}[.]{Reichenbach, Carnap, Hempel, and Sellars gave principled reasons why a finite translation of material-thing language into sense-datum language was impossible. 
Even if these reasons fall short of a strict mathematical impossibility proof, they are enormously convincing [\ldots]. In the same spirit, I am going to give principled reasons why a finite empirical definition of intentional relations and properties in terms of physical/computational relations and properties is impossible---reasons which fall short of a strict proof, but which are, I believe, nevertheless convincing}
 }
 \section{Conclusion}\label{sec:conclusion}
\noindent
Straight solutions to the measurement problem suggest that one has to choose among the available ``interpretations'' of quantum mechanics.
The common way of putting the measurement problem is, however, itself problematic:
\emph{It relies on reducing science to pure description short of any normative elements.}
In particular, describing a measurement \emph{within} a theory is faced by the problematic traversing from the universal to the particular---and back:
How can reference to particular systems be drawn from a general framework?
\emph{How do particular experiences warrant universal truths?}
More generally, how is linguistic description possible without establishing intersubjective agreement?
We do not call for \emph{definite} answers to these questions.
Any such answer, we suspect, creates its own contingencies.
\emph{We conclude, however, that it is up to interpretation whether there is a problem in the first place.}
We ask further, supposing that there \emph{is} a problem, whether it is the one that is exposed in the common framing.
Instead of inviting to choose ones favorite ``interpretation,'' the ``problem'' \emph{asks to reflect} on what ``doing physics'' means.
 
\hbox{\ }
\begin{acknowledgments}
\noindent
  This work is supported by the Swiss National Science Foundation (SNF), the \emph{NCCR QSIT}, and the \emph{Hasler Foundation}. We would like to thank \"Amin Baumeler, Veronika Baumann, Cecilia Boschini, Paul Erker, Claus Beisbart, Xavier Coiteux-Roy, Manuel Gil, and Christian W\"{u}thrich for helpful discussions.
\end{acknowledgments}

\end{document}